\documentstyle[12pt,a4]{article}
\begin{document}
\newcommand{\wst}{~{}^{{}^{*}}\llap{$\it w$}}
\newcommand{\wdst}{~~~{}^{{}^{**}}\llap{$\it w$}}
\newcommand{\omegast}{~{}^{{}^{*}}\llap{$\omega$}}
\newcommand{\omegadst}{~~~{}^{{}^{**}}\llap{$\omega$}}
\newcommand{\must}{~{}^{{}^{*}}\llap{$\mu$}}
\newcommand{\mudst}{~{}^{{}^{**}}\llap{$\mu$~}}
\newcommand{\nust}{~{}^{{}^{*}}\llap{$\nu$}}
\newcommand{\nudst}{~{}^{{}^{**}}\llap{$\nu$~}}
\newcommand{\beq}{\begin{equation}}
\newcommand{\eeq}{\end{equation}}
\newcommand{\gfrc}[2]{\mbox{$ {\textstyle{\frac{#1}{#2} }\displaystyle}$}} 
\title {\large \bf Higher Dimensional  Cosmological Model in \\ 
Space-Time-Mass (STM)  Theory of Gravitation}
\author{ G S Khadekar \thanks{Tel.91-0712-23946,
 email:gkhadekar@yahoo.com} and Shilpa Samdurkar \\
Department of Mathematics, Nagpur University \\Mahatma Jyotiba Phule 
Educational Campus,  \\ Nagpur-440 010 (INDIA) }
\maketitle
\begin{abstract}
A new class of non-static higher dimensional vacuum solutions in 
space-time -mass (STM) theory of gravity is found. This solution 
represent expanding universe without  big bang singularity and the 
higher dimension of these models shrinks as they expands.
\end{abstract}
\section{ Introduction}
 The domination of Newtonian theory had received the blow at 
the advent of Einstein general theory of relativity, where the 
Newtonian gravitational parameter G resticted as constant. Several 
theories of gravitation ( Brans  and Dicke [1], Dirac[2], Hoyle 
Narlikar[3] and Canuto et al.[4] ) alternatives to Einstein 
gereral theory of relativity have been proposed on which the 
parameter G and or mass m vary slowly with time.These theories have the common feature that they assume the absences of fixed scales in cosmological solutions over times comparable to the age of the universe. \par  In this contents, the variable gravitational theory proposed by Wesson [5,  6] deserved serious attention. The space-time five dimension with coordinate $ x^0 = ct $ ( c=velocity of light, t = time), $ x^{1,2,3} $ (space coordinate) and $ x^4 =\frac{Gm}{c^2} ( m= mass).$  The  Einstein theory is recovered when  the velocity  $ \: \frac{Gm}{c^2} \gfrc{dm}{dt} = 0 .$ In some sense, the usual  Einstein theory Einstein theory would be embedded in it.This new five dimensional theory of variable rest mass is a nutural extension a of relativity theory. In our work, we have extend the work of L K Chi [7] for n-dimensional variable mass theory of gravity. Here we intend to investigate the new class of vacuum solutions in n-dimensional space-time-mass (STM) theory of gravity. These cosmological higher dimensional models describe expanding universe without a big bang singularity and the higher dimensional of these models shrinks as they expand.      
\subsection{Field Equations}
 The line element for a n-dimensional  homogeneous and
spatially isotropic  cosmological model is taken as \\
\begin{equation}
ds^2 = e^\nu dt^2-e^{\omega}\sum_{i=1}^{(n-2)}dx_{i}^2 + e^{\mu} dm^2
\end{equation}
where $ \mu,{\omega} $ and $\nu $ are the functions of time
and mass.Here the coordinate $x^0 = t, \: x^{1,2, \cdots, (n-2) }$
(space coordinate) and $x^{(n-1)} = m.$ For simlicity we have set the
magnitudes of both c and G to unity. By applying this metric to the
Einstein field equation $ G_{ij} = R_{ij} -\gfrc{1}{2} g_{ij} \, R =0,
\; \; $  we get
\begin{equation}
G_{00} = -(n-2)(n-3) \; \frac{\dot \omega^2}{8}- (n-2) \: \frac{\dot \omega\dot \mu}{4} - (n-2) \, e^{\nu-\mu}\left( \frac{\omegadst}{2}-\frac{\omegast \must}{4} + (n-1) \, \frac{\omegast^2 }{8}  \right)
\end{equation}
 \begin{eqnarray}
G_{11} = e^{\omega-\mu} \left(\frac{\ddot \mu}{2}  + \frac{\dot \mu^2}{4} -\frac{\dot {\mu }\dot {\nu}}{4}\right) +  (n-3)e^{\omega-\nu}  \left(\frac{\ddot \omega }{2} +\frac{\dot{ \omega} \dot {\mu}}{4} - \frac{\dot \nu \dot \omega}{4} +(n-2)   
\frac{{\ddot \omega^2} }{8} \right)    \nonumber \\ +  (n-3)  e^{\omega- \mu} \left( \frac{\omegadst }{2} + \frac{ \omegast \must}{4} - \frac{ \nust  \omegast}{4} +  (n-2)  \frac{ \omegast^2}{8} \right)  +  e^{\omega - \mu} \left(\frac{\nudst}{2}  + \frac{ \nust^2}{4} -\frac{\nust \must}{4}\right)   
\end{eqnarray}
 $$ G_{11} = G_{22} = G_{33} = \cdots = G_{(n-2)(n-2)} $$
\beq
G_{0(n-1)} = (n-2) \left( \frac{\dot{\omega}^*}{2 }+ \frac{\omegast \: \dot{ \omega}}{4}- \frac{\nust  \: \dot \omega}{4} -\frac{\dot \mu \omegast}{4}\right)
\eeq
\beq
G_{(n-1)(n-1)} = -(n-2)(n-3) \frac{\omegast^2}{8}- (n-2) \frac{\omegast \nust}{4} - (n-2) \;  e^{\mu-\nu}\left(\frac{ \ddot \omega}{2} -\frac{\dot {\omega}\dot{\nu}}{4} +(n-1)\frac{\dot{\omega^2}}{8}\right) 
\eeq
 where  a dot and star denote, respectively partial derivative with respect to time and mass.
\subsection{Solutions}
Equationa (2) , (3) and (5) have the peculiar property that the time derivative and mass derivative are completly separeted. \\ Hence we get 
\beq
(n-3) \: \frac{\dot{\omega^2}}{2} +\dot{\omega}\dot{\mu} = 0
\eeq
\beq
(n-3) \; \frac{\omegast^2}{2} +\omegast \nust = 0
\eeq
\beq
\omegadst -\frac{\omegast \must}{2} + (n-1) \: \frac{\omegast^2}{4} = 0
\eeq
\beq
\ddot{\omega} - \frac{\dot {\omega} \dot{\nu}}{2} +(n-1) \: \frac{\dot{\omega}^2}{4} = 0
\eeq
assuming that $ \omegast  $ and $ \dot{\omega} $ are not zero, it then follows from equations (6) and (7) 
\beq
\dot{\omega} = - \frac{2\: \dot{\mu}}{(n-3)}
\eeq
\beq
\omegast = - \frac{2\; \nust}{(n-3)}
\eeq
with the aid of equations (8) - (11), we find that equation (3) is identically satisfied. \\ Using equations (10) and (11), we can integrate equation (4) to give 
\beq
e^{\omega} = \left[ \frac{(n-2)}{2} \int k(m) \: dm + r(t) \right]^{\frac{2}{(n-2)}} 
\eeq  
where $ k(m) $ is an arbitary function of m and r(t) is an arbitary function of t. once $ \omega $ is known, $\mu $ and $ \nu $ can easily be found from equations (10) and (11).
\beq
e^{\mu} = \frac{ P\: k(m)^2}{\left[ \frac{(n-2)}{2} \int k(m) \: dm + r(t) \right]^{\frac{n-3}{(n-2)}}}
\eeq
\beq
e^{\nu} = \frac{ Q \: r(t)^2}{\left[ \frac{(n-2)}{2} \int k(m) \: dm + r(t) \right]^{\frac{2}{(n-2)}}}
\eeq
wherw P and Q are integrating constant.
\subsection{Conclusion}
\par The implification of the condition $ \dot{\omega}  = 0 $ is that the function  r(t)  can not be a constant and these cosmological models are not static. Equation (10) implies that the higher dimensions shrinks as the universe expand. This conclusion can also be drawn from the explicit expansion (12) and (13). It it interesting to note that if  r(t)  is an increasing function of time t  and the function  k(m)  is bounded, then $ e^{\mu} $ becomes smaller as $ e^{\omega} $ increase. For the expanding universe, we may choose the function  r(t)  to be positive function of t or an exponential function of t. The condition $ \omegast \ne 0  $ implies that we can not choose the arbitary fubnction  k(m)  to be zero. Hence, this expanding universe in higher dimensional STM theory donot start with a big bang singularity.
\bibliographystyle{plain}

\end{document}